\documentclass[apl,twocolumn,showpacs,amsmath,letter]{revtex4}
\usepackage{graphicx}

\newcommand{\Journal}[4]{#1 \textbf{#2}, #3 (#4)}

\begin{document}

\title{Studies of Current-Driven Excitations in Co/Cu/Co Trilayer
Nanopillars}

\author{S. Urazhdin, Norman O. Birge, W.P. Pratt, Jr., J. Bass}
%\author{N. O. Birge}
%\author{W. P. Pratt Jr.}
%\author{J. Bass}
\affiliation{Dept. of Physics and Astronomy, Center for
Fundamental Materials Research, and Center for Sensor Materials,
Michigan State University, East Lansing, MI 48824-2320}


\begin{abstract}
We measure the dynamic resistance of a Co/Cu/Co trilayer
nanopillar at varied magnetic field $H$ and current $I$. The
resistance displays the usual behavior, almost symmetric in $H$,
both when magnetization switching is hysteretic at small $I,H$,
and reversible at larger $I,H$. We show differences in the $I,H$
magnetization stability diagram measured by holding $I$ fixed and
varying $H$ and vice versa. We also show how the peak in $dV/dI$
associated with telegraph noise in the reversible switching
regime, is calculated from the telegraph noise variations with
$I$. Lastly, we show data for a similar sample that displays
behavior asymmetric in $H$, and a negative reversible switching
peak instead of a usual positive one.
\end{abstract}

\maketitle

Current-induced switching of magnetization has generated much
excitement due to its potential for magnetic random access memory.
In spite of the apparent success of the spin-torque
model~\cite{slonczewski} in describing many of the experiments,
the basic physical processes involved in the switching are not yet
fully understood. Most experimental studies of current-driven
magnetization switching in magnetic nanopillars have been made on
Co/Cu/Co trilayers at room temperature
(295~K)~\cite{cornellapl,cornellorig,cornellquant,cornelltemp,mancoff,sun2,grollier2,kent,urazhdinAPL}.
For magnetically uncoupled samples, switching at low current $I$
and magnetic field $H$ is hysteretic, but becomes reversible at
large enough $I$ in one direction. This reversibility is
associated with telegraph noise switching~\cite{urazhdinPRL}. In
this paper we examine several subtleties of switching in Co/Cu/Co
that have not been previously described. First, the $I$ {\it vs.}
$H$ switching (magnetization stability) diagrams are slightly
different when measured by varying $H$ while holding $I$ fixed and
vice versa. Second, we show how the reversible switching peak can
be calculated from the measurements of telegraph noise dwell times
{\it vs.} $I$. Third, we show data for an unusual sample, where a
positive reversible switching peak is replaced by a negative one.

Our samples were fabricated with a multistep process described
elsewhere~\cite{urazhdinAPL}. The samples had structure
Co(20)/Cu(10)/Co(2.5), where thicknesses are in nm. To minimize
dipolar coupling between the Co layers, only the Cu(10)/Co(2.5)
layers were patterned into a nanopillar with approximate
dimensions $140\times 70$~nm. We measured differential resistance,
$dV/dI$, at 295~K with four-probes and lock-in detection, adding
an ac current of amplitude 20~$\mu$A at 8~kHz to the dc current
$I$. Positive current flows from the extended to the patterned Co
layer. $H$ is in the film plane along the easy axis of the
nanopillar.

\begin{figure}
\centering
\includegraphics[scale=0.4]{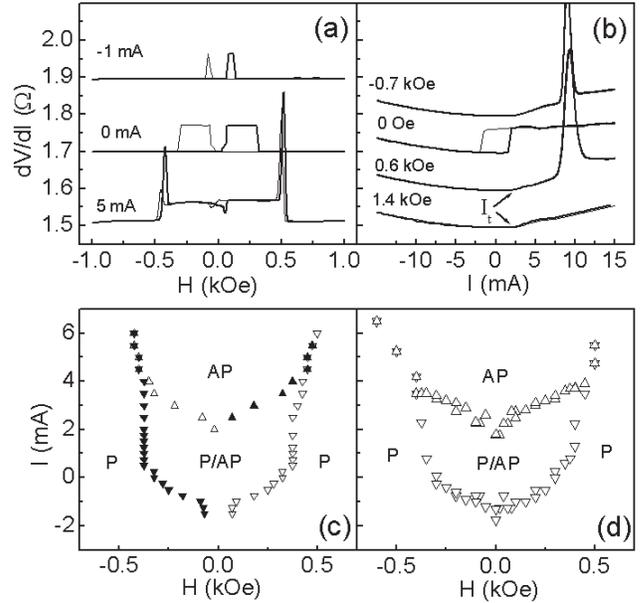}
\caption{\label{fig1} Data for sample 1. (a) $H$-dependence of
$dV/dI$ at specified values of $I$; (b) $I$-dependence of $dV/dI$
at specified values of $H$. $I_t$ denotes the excitation threshold
current. In (a),(b), thick curves: scan from left to right, thin
curves: scan in opposite direction, and curves are offset for
clarity. (c) Magnetization stability diagram extracted from $H$
scans such as shown in (a). Upward(downward) triangles:
P$\to$AP(AP$\to$P) switching. Open symbols: scan from left to
right, solid symbols: reverse scan direction. (d) Magnetization
stability diagram extracted from $I$ scans such as shown in (b).
Upward(downward) triangles: P$\to$AP(AP$\to$P) switching. In
(c),(d) AP, P denote the stability region of the respective
configurations, P/AP is a bistable region.}
\end{figure}

Figs.~\ref{fig1}(a,b) show field- and current-switching data,
consistent with prior
studies~\cite{cornellapl,urazhdinAPL,urazhdinPRL}. Starting, for
example, at $I=0$ and large negative $H$, the magnetizations of
the thick and thin Co layers are parallel (P). As $H$ is increased
from a large negative value, the magnetization of Co(20) switches
first at small positive $H$ into a high resistance antiparallel
(AP) state, and the patterned Co(2.5) switches at larger switching
field, $H_s(I=0)$, determined by its shape anisotropy. For
$I=-1$~mA, Fig.~\ref{fig1}(a) shows reduced $H_s(I)$, and the
hysteretic switching disappears at $I<-1$~mA. $I>0$ increases the
range of $H$ for the AP configuration. At $I>4$~mA, the hysteretic
switching steps in $dV/dI$ turn into reversible peaks ($I=5$~mA
shown). Fig.~\ref{fig1}(b) shows that hysteretic asymmetric
current-driven switching between the AP state at $I>0$ and the P
state at $I<0$ at $H=0$, changes to reversible peaks both at large
$H>0$ and $H<0$. These peaks are the same as those in
Fig.~\ref{fig1}(a) at large $I>0$. The P state resistance grows
above a threshold $I_t$, marked on the $H=0.6,1.4$~kOe curves. A
similar, more pronounced threshold in Py/Cu/Py nanopillars has
been associated with the onset of large amplitude magnetic
excitations~\cite{urazhdinPRL}. At small $H$, the switching from P
to AP state occurs at $I_s\approx I_t$. The small variation of
$I_t$ between $0.6$~kOe and $1.4$~kOe in Fig.~\ref{fig1}(b) is
determined by the balance between the current-driven excitation
and weakly $H$-dependent magnetic damping rate.

Figs.~\ref{fig1}(c,d) show the Co(2.5) nanopillar magnetization
stability diagrams extracted from $H$ and $I$ scans such as those
in Figs.~\ref{fig1}(a,b), respectively. (We show only the
switching of the thin Co layer in Fig.~\ref{fig1}(c), to avoid
clutter and facilitate comparison with Fig.~\ref{fig1}(d).) Both
scan directions give similar stability regions, with a minor
difference in the line separating the bistable and P-stable
regions. At small $I>0$, the stability line in Fig.~\ref{fig1}(c)
is almost vertical, giving a sharp knee at $I=0$, whereas in
Fig.~\ref{fig1}(d) it curves smoothly at $I\approx 0$. Vertical
lines are poorly reproduced by $I$-scans, so Fig.~\ref{fig1}(c)
better captures the singular behavior at $I\approx 0$. This knee
has been attributed to the effect of spontaneous current-driven
magnon emission, generally small compared to stimulated
emission~\cite{urazhdintheory}.

\begin{figure}
\centering
\includegraphics[scale=0.4]{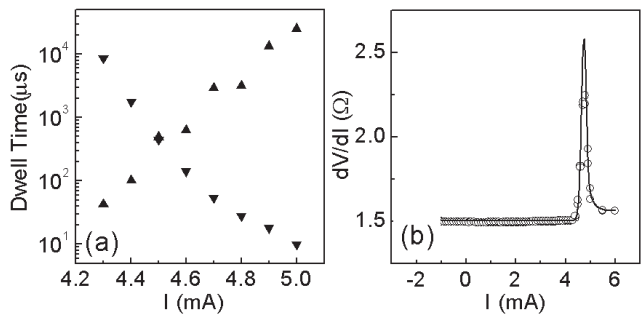}
\caption{\label{fig2} Data for sample 1. (a) Variation of the
average dwell times in the AP state $\tau_{AP}$ (upward triangles)
and P state $\tau_P$ (downward triangles) with $I$ at
$H=0.48$~kOe, (b) Circles: $dV/dI$ {\itshape vs.} $I$ at
$H=0.5$~kOe. Solid curve: a calculation, as described in the
text.}
\end{figure}

The reversible switching peaks in $dV/dI$ that at large $I$,$H$
replace the hysteretic steps, are due to random telegraph noise
switching between the P and AP states~\cite{urazhdinPRL}.
Fig.~\ref{fig2}(a) shows the variations of average dwell times
$\tau_P(\tau_{AP}$)in the P(AP) state with $I$. $\tau_P$ decreases
as $I$ increases, but $\tau_{AP}$ increases. For a fixed $H$,
$\tau_{AP}<<\tau_{P}$ at small $I$, so $dV/dI$ is close to the
resistance of the P state, $R_{P}$, and $\tau_{AP}>>\tau_{P}$ at
large $I$, giving $dV/dI\approx R_{AP}$, the resistance in the AP
state. We now show how the variations in Fig.~\ref{fig2}(a) give a
peak in the differential resistance at $\tau_{P}\approx\tau_{AP}$.
For a given $H$, the average voltage across the sample is
\begin{equation}\label{voltage}
V(I)=I\left[\frac{R_{AP}\tau_{AP}+R_{P}\tau_{P}}{\tau_{P}+\tau_{AP}}\right],
\end{equation}
where $\tau_{P}(I)\approx\tau_0\exp[-\alpha(I-I_0)]$,
$\tau_{AP}(I)\approx\tau_0\exp[\beta(I-I_0)]$, as follows from
Fig.~\ref{fig2}(a). $I_0$, $\tau_0$ are defined by
$\tau_{AP}(I_0)=\tau_{P}(I_0)=\tau_0$. Differentiating
Eq.~\ref{voltage} with respect to $I$, we find

\begin{eqnarray}\label{dvdi}
\frac{dV}{dI}\approx
\nonumber\frac{\tau_{AP}R_{AP}+\tau_{P}R_P}{\tau_{P}+\tau_{AP}}+\\
I(\alpha+\beta)(R_{AP}-R_P)\frac{\tau_{P}\tau_{AP}}{(\tau_{P}+\tau_{AP})^2}.
\end{eqnarray}

The first term on the right is just the resistance $V/I$, giving a
step for the reversible transition from the P to the AP state. The
second term has a maximum value
$I_0(\alpha+\beta)(R_{AP}-R_{R})/4$ at $\tau_{P}=\tau_{AP}$. This
term gives rise to a peak in $dV/dI$ at $I=I_0$, which can be much
higher than $R_{AP}$. The solid line in Fig.~\ref{fig2}(b) is
calculated from the data in Fig.~\ref{fig2}(a), and
Eq.~\ref{dvdi}, for $I_0=4.8$~mA, and
$\alpha+\beta=19.2$~mA$^{-1}$ extracted from fig.~\ref{fig2}(a).
The calculation agrees well with the data shown as circles.

>From the above analysis, we conclude that the reversible switching
peak positions characterize the points $(H,I)$ where
$\tau_P=\tau_{AP}$, thus giving an indirect measure for telegraph
noise variation with $I,H$~\cite{largeIHnote}. We have
shown~\cite{urazhdinPRL} that the telegraph noise period decreases
approximately exponentially when $I$ is increased and $H$ is
adjusted to remain along the reversible switching line. The
presence of telegraph noise near the reversible switching line
means that both AP and P states are unstable in that region. Thus,
the stability diagrams, Figs.~\ref{fig1}(c,d), should be modified
to include this unstable region. This instability is indirectly
manifested in the rise of $R_P$ at $I>I_t$. However, the
measurements of $dV/dI$ at $I$ above the reversible switching peak
give values very close to $R_{AP}$. Fig.~\ref{fig2} and our
analysis show that, because $\tau_{P}$ is exponentially smaller
than $\tau_{AP}$, the resistance can become close to $R_{AP}$,
even though the AP state is unstable.

\begin{figure}
\centering
\includegraphics[scale=0.4]{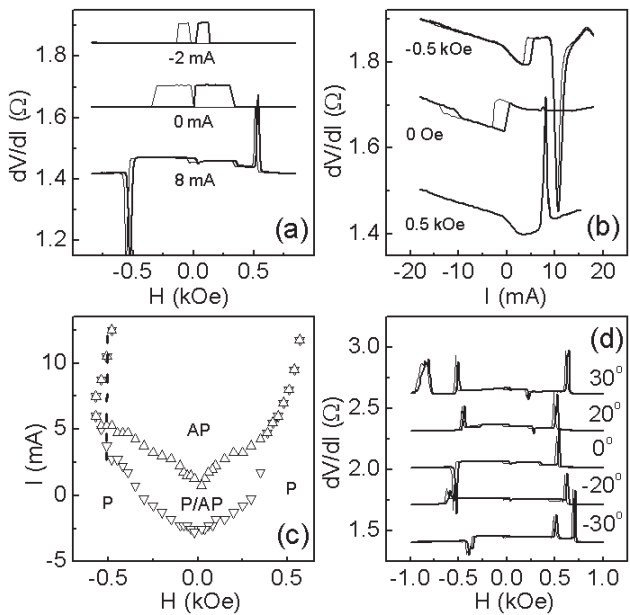} \caption{\label{fig3}
Data for sample 2. (a) $H$-dependence of $dV/dI$ at specified
values of $I$, (b) $I$-dependence of $dV/dI$ at specified values
of $H$. (c) Magnetization stability diagram extracted from $I$
scans such as shown in (b). Upward(downward) triangles:
P$\to$AP(AP$\to$P) switching. A $H=-0.5$~kOe section shown with
dashed line, (d) MR curves at $I=8$~mA, at the specified in-plane
angles between $H$ and the nominal easy axis of the nanopillar.
AP, P denote the stability region of the respective
configurations, P/AP is a bistable region. In (a),(b),(d), thick
curves: scan from left to right, thin curves: scan in opposite
direction, curves are offset for clarity.}
\end{figure}

In most samples, both $I$ and $|H|$ increase along the reversible
switching line close to the transition from hysteretic to
reversible switching. The behavior at larger $I$ varies: in some
samples, the reversible switching peak disappears, or splits into
several peaks. These peaks are usually asymmetric in $H$, showing
the importance of inhomogeneous  and tilted magnetization states,
affected both by sample imperfections and the Oersted field of the
current. Fig.~\ref{fig3} shows data for a sample nominally
identical to that of Fig.~\ref{fig1}. The hysteretic MR at $I=0$,
and current-driven switching at $H=0$ (Figs.~\ref{fig3}(a,b)), are
similar to those in Figs.~\ref{fig1}(a,b). The $I=8$~mA MR curve
in Fig.~\ref{fig3}(a) is asymmetric, showing a positive peak at
$H>0$ like those in the $5$~mA curve in Fig.~\ref{fig1}(a), but a
negative peak at $H<0$. Similarly, in the current scans of
Fig.~\ref{fig3}(b), the peak at $H=0.5$~kOe is consistent with
those at $-0.7,0.6$~kOe in Fig.~\ref{fig1}(b), while the
$-0.5$~kOe scan shows a small hysteresis in current switching with
a negative peak at larger $I$. By comparing the $8$~mA resistances
to the left of the negative peak and to the right of the positive
peak in Fig.~\ref{fig3}(a), we conclude that the negative peak
corresponds to complete AP$\to$P switching. The resistance
increase to the right of the negative peak in Fig.~\ref{fig3}(b)
is consistent with the previously noted current-driven excitations
in the P state~\cite{urazhdinPRL}. Fig.~\ref{fig3}(c) shows the
stability diagrams extracted from $I$ scans such as those in
Fig.~\ref{fig3}(b), where we mark both the positive and negative
peaks as reversible switching points. This plot clearly shows the
asymmetry of behaviors with respect to reversal of $H$. For $H>0$
the stability diagram is similar to those of
Figs.~\ref{fig1}(c,d). For $H<0$ in Fig.~\ref{fig3}(c), the
reversible switching line has a positive slope, i.e. the negative
peaks appear at decreasing $I$ as the magnitude of $H$ is
increased. A dashed $H=-0.5$~kOe line crosses both a bistable
region (hysteretic switching), and a reversible switching line.
The positive slope of the reversible line is consistent with
$\alpha+\beta<0$, giving a negative peak in Eq.~\ref{dvdi}.

Fig.~\ref{fig3}(d) shows $H$-scans at $I=8$~mA with varied angles
$\theta$ between the nominal easy nanopillar axis and $H$ directed
in the sample plane. The $\theta=0$ curve has a positive peak at
$H>0$ and negative peak at $H<0$. The peaks in the $\theta=\pm
20^o$ curves are nearly symmetric, and positive for both
directions of $H$. The $\theta=\pm 30^o$ curves are asymmetric
again, and have double peaks for one of $H$ directions. These data
show that the details of switching are sensitive to the sample
shape defects, misalignment of the nanopillar easy axis with $H$,
and are also affected by the Oersted field of the current and
magnetization pinning. We note that only the last two factors
(possibly in combination with the first two) give asymmetry
between the behaviors at $H<0$ and $H>0$.

To summarize, we focused on four phenomena in Co/Cu/Co nanopillars
at 295~K. First, we provided evidence (although not as clear as in
Py/Cu/Py~\cite{urazhdinPRL}) of a threshold current $I_t$ for
excitations that occur in the reversible switching regime, but at
lower $I$ than the reversible switching peak. Second, we showed
that the sharp knee at $I=0$, visible in a magnetization switching
diagram obtained by fixing $I$ and varying $H$, is lost in a
similar plot obtained by fixing $H$ and varying $I$. Third, we
showed that the reversible switching peak shape can be derived
from measurements of the variation of telegraph noise with $I$.
Fourth, in Fig.~\ref{fig3} we showed an example of a switching
diagram asymmetric in $H$, more complex than the symmetric one in
Fig.~\ref{fig1}. We attribute the complexity to a combination of
sample shape asymmetry, the Oersted field, and possible
misalignment of $H$. Of particular interest in Fig.~\ref{fig3}(b)
is the negative peak, associated with re-entrance of the P state
at high $I>0$.

We acknowledge support from the MSU CFMR, CSM, the MSU Keck
Microfabrication facility, the NSF through Grants DMR 02-02476,
NSF-EU 98-09688, and NSF-EU 00-98803, and Seagate Technology.


\begin{thebibliography}{99}
\bibitem{largeIHnote} If we follow the reversible switching peak to very large $I$ and $H$, the dwell times $\tau_{P}$ and $\tau_{P}$ extrapolate to values
smaller than any plausible microscopic attempt time. At that point
the magnetization dynamics no longer produce telegraph noise, but
become more complicated.
\end{thebibliography}
\end{document}